\documentclass[12pt,a4paper]{article}

\usepackage[margin=3cm]{geometry}

\usepackage{authblk}
\usepackage{bm}
\usepackage{booktabs}
\usepackage{hyperref}
\usepackage{lmodern}
\usepackage[numbers]{natbib}

\usepackage{amsmath}
\usepackage{cleveref}

\usepackage{epstopdf}
\usepackage{graphicx}

\newcommand{\tl}{\texttt{TubeLab}}

\begin{document}

\title{\tl{}: Interactive inverse design of wind instrument bores with hard spectral constraints}
\author[1]{Carlo Andrea Rozzi}
\affil[1]{Istituto Nanoscienze –- CNR, S3, via Campi 213/A, 41125 Modena, Italy}
\author[2]{Andrea Ferroni}
\affil[2]{via Regione Montebruno 2/A, 10060 Garzigliana (TO), Italy}
\date{\today}

\begin{abstract}
We present \tl{}, a browser-based acoustic simulator and inverse design tool for wind instrument bores without tone holes. The direct problem is solved by a Transfer Matrix Method. Inverse design — finding a bore correction that shifts selected resonance frequencies toward specified musical targets — is formulated as a Tikhonov-regularized least-squares problem whose Jacobian is computed by a single adjoint backward pass per mode. Modes whose frequencies must be preserved are handled by a saddle-point formulation that enforces hard equality constraints regardless of the regularization strength. Resonance frequencies agree with state-of-the-art finite-element calculations to within 1.21 cent over eleven geometries and 132 mode pairs. Experimental validation against tap-tone measurements of 16 numerical control-machined didgeridoos spanning six distinct bore designs yields 9.6 cent RMS deviation over 189 matched mode pairs. Replicate specimens of the same nominal bore disagree with each other by 4.5 cent RMS, so about 36\% of the residual variance is fabrication scatter.
\end{abstract}

\maketitle

\section{Introduction}
\label{sec:intro}

The traditional path of instrument making moves from materials to shape to sound. For centuries empirical working experience, guild transmission and player feedback have informed makers' design choices. In the pre-scientific era no maker is documented to have worked backward from an acoustic target to a geometry in any formal sense. Most instruments just evolved because materials and manufacturing techniques changed over time and makers' experience accumulated and adapted accordingly.

The earliest known case of a maker reasoning acoustically about bore shape dates back to the early 18th century, when Johann C. Denner realized that a cylindrical bore overblows the twelfth (not the octave). This observation was exploited to differentiate the clarinet register structure from the chalumeau~\cite{Karp1986}. The clearest historical case of designing from acoustic targets is the development of the modern flute made by Theobald B\"ohm in the 19th century. However this is still forward design (applying theory to choose geometry). As a matter of fact heuristically finding solutions to the inverse problem (designing a geometry to match an arbitrary acoustic response) turns out to be much harder.

Mathematicians and physicists have been facing inversion problems in many contexts since long time~\cite{Ambarzumian1929}, but the mathematical problem of iso-spectral domains certainly acquired widespread popularity as soon as it was phrased in terms of drum skins~\cite{Kac1966}. Today we know that there is no unique shape corresponding to a given drum sound~\cite{Buser1994}. However the one-dimensional problem -- sound propagation in a duct -- does admit a unique inverse solution at least in the case the full complex impedance across all frequencies is known. In this case the shape of the tube that produced it can be fully reconstructed. This works ideally. In practice all measurements are band-limited and noise-polluted, making the inversion problem ill-posed, requiring some form of regularisation~\cite{SondhiGopinath1971}. Regardless technical details, reconstructing bores from laboratory full acoustic measurements (impedance or acoustic pulse reflectometry) has become an established technique in wind instruments research~\cite{SondhiResnick1983, Bowen201984, Forbes2006, Hendrie2007, Ernoult2021}.

A complementary strand of research addresses the inverse design problem: finding a bore shape that realizes a prescribed spectral target. \citet{Kausel2001} pioneered the systematic optimization of brass instrument bores using a gradient-free algorithm with up to 100 geometric degrees of freedom, showing that computer search over a parametrized bore space can achieve specified intonation targets. \citet{Noreland2010} introduced a gradient-based approach in which the analytic sensitivity of resonance frequencies to bore perturbations is computed via a hybrid transmission matrix method (TMM) / finite-element model (FEM), demonstrating that large parameter counts become tractable once the gradient is available analytically. \citet{Braden2009} applied bore optimization directly against target input impedance curves derived from reference trombones; \citet{Macaluso2011} showed that optimized geometries can be physically realized, building a trumpet whose resonances deviate from perfect harmonicity by only 5 cent RMS. The most complete treatment of the problem for woodwind instruments with tone holes is due to \citet{Ernoult2020}, who derived adjoint-state gradients of both resonance frequencies and modal amplitudes from the TMM, and used them to optimise a keyless pentatonic clarinet subject to geometric manufacturing constraints. OpenWinD~\cite{Chabassier2020OpenWinD} further consolidates this body of work into a freely available Python library and online simulation tool combining forward impedance computation, bore reconstruction by full-waveform inversion, and impedance-targeted optimization. These contributions demonstrate that the computational design of wind instrument bores from spectral targets is feasible and can yield acoustically superior geometries; however, in all these works the optimization is posed as a fully automated, expert-facing numerical procedure.

Can this knowledge be transferred to the practice of instrument makers? Makers usually do not have professional apparatuses to make full accurate impedance measurements, and reconstructing a bore shape after it is built is not a common necessity. They might still be very interested in an acoustically informed framework to tune the overtones and to explore several compatible shapes for playability and easiness of manufacturing before starting to build a new instrument. They might also want to amend or reshape an instrument to change its timbre via retuning a few selected partials. The knowledge of a discrete set of resonance frequencies alone (instead of the full input impedance) does not uniquely determine the bore, but this freedom can be exploited as it offers greater possibilities to manufacturers, which are, on the other hand, often bound to physical constraints about what actions can or can't be realistically done to modify the bore. In all cases, a physically grounded framework is necessary, but easiness of use, adherence to realistic workflows and interactivity are key factors too.

We addressed the simplest case of bores with no tone-holes. In this category fall, for example, natural trumpets and horns. However, while the bores of these instruments are traditionally limited to a few canonical shapes (conical or cylindrical, possibly with a bell flare), the didgeridoo eludes all standardizations. 

Didgeridoo is the common name for a traditional instrument of the Aboriginal people from North Australia. It is historically obtained from Eucalyptus trees hollowed by termites, yielding irregular inner cavities. Modern, non ritual design of the didgeridoo allows for more varied (possibly curved) external shapes and more variety in the material choice, including bamboo, fiberglass, PVC, resin and metal, besides several kinds of wood, which are usually carved starting from flat beams, and then joined. The bore is more than 1~m long and is played as a brass instrument, i.e., by vibrating lips directly into it. Lip action is often accompanied by circular breathing and tongue rhythmic effects. Besides the fundamental frequency, players can also excite a few overtones by adjusting lip tension. Moreover the players' oral cavity couples to the bore thanks to its low input impedance to greatly enrich the sound color palette~\cite{Tarnopolsky2005}.

The freedom in length, size and shape provides great variability in the timbre and playability of the bores, but also leaves makers the option either to resort to known templates, or to manage a great deal of trial and error exploration in case they are searching for specific acoustic qualities.

Here we propose a simple, computationally efficient way to solve the direct (\cref{sec:physics}) and inverse problem (\cref{sec:inverse}) for didgeridoos starting from musically meaningful overtone tuning targets, subject to realistic constraints a maker might want (or be forced to) fulfill. We embedded a novel physics-informed workflow in a browser-native web app (\tl{}) allowing makers to interactively explore and adjust shapes. We validated the app by comparing its predictions to state of the art FEM software and to experimental impulse response measurements performed on a set of instruments fabricated with a numerically controlled (NC) milling machine (\cref{sec:validation}). The app also has valuable educational features. As such it is included in a publicly accessible repository of didactic material about acoustics and waves hosted by the University of Modena and Reggio Emilia~\cite{FOM2006}.

\section{Bore acoustic model}
\label{sec:physics}

The direct problem (from geometry to sound spectrum) is formulated within the classical TMM, which solves the Webster's horn equation~\cite{Webster1919} in the frequency domain to calculate the bore's input impedance. In \tl{} the bore shape can be conveniently manipulated by the user by moving a small number of nodal control points (optionally interpolated to a smooth line via a monotone cubic Hermite spline~\cite{FritschCarlson1980}). However, for acoustic modeling it is discretized over a regular grid of points $\{x_i, r_i\}$ with $i \in [1, N]$, $N=200$. Typical didgeridoos shapes do not have extreme bell flares, unlike other brasses, and the piecewise-cylindrical approximation of the elements does not impact results (more details in \cref{sec:validation}). The area function of an element $\bar S_i$ is the average of its input $i$ and output $i+1$ areas. For each segment the acoustic pressure, volume velocity state vector $(p, U)$ obeys the transfer relation~\cite{ChaigneKergomard2016}
\begin{equation}
  \begin{pmatrix} p_i \\ U_i \end{pmatrix}
  = \begin{pmatrix} \cos k\Delta x & \mathrm{i} Z_c \sin k\Delta x \\
    \mathrm{i} Z_c^{-1} \sin k\Delta x & \cos k\Delta x \end{pmatrix}
  \begin{pmatrix} p_{i+1} \\ U_{i+1} \end{pmatrix},
  \label{eq:TM}
\end{equation}
where $Z_c = \rho c / \bar S_i$ is the characteristic impedance of the segment and $k$ is the complex wave number (see below).

The input impedance $Z_\text{in}(f)$ is computed by M\"obius recursion backwards from the bell at 1800 equally-spaced frequency points covering $[20, 5000]$\,Hz. The recursion performs $O(N)$ complex arithmetic operations per frequency point and is numerically stable because the Möbius transformation preserves the positive-real character of the impedance.

For didgeridoos the mouth is always treated as a rigid termination, that is a pressure antinode, while at the bell the \citet{LevineSchwinger1948} unflanged radiation is imposed through the Pad\'e approximation of \citet{Silva2009}, which reproduces the exact Wiener--Hopf solution to better than $2\,\%$ for $kr_\text{bell} < 3$. Writing $x = kr_\text{bell}$, the reflection modulus $|\mathcal{R}|$ and the end correction $\ell$ are fitted separately,
\begin{equation}
  \begin{split}
  |\mathcal{R}| &= \frac{1 + a_1x^2}{1 + (\beta + a_1)x^2 + a_2x^4 + a_3x^6}, \\[2pt]
  \frac{\ell}{r_\text{bell}} &= \eta\,\frac{1 + b_1x^2}{1 + b_2x^2 + b_3x^4 + b_4x^6},
  \end{split}
  \label{eq:radimpedance_unflanged}
\end{equation}
with $\beta = 1/2$, $\eta = 0.6133$, $a_1 = 0.800$, $a_2 = 0.266$, $a_3 = 0.0263$, $b_1 = 0.0599$, $b_2 = 0.238$, $b_3 = -0.0153$ and $b_4 = 0.00150$; the impedance follows as 
\begin{equation}
  \frac{Z_\text{rad}}{Z_{c,\text{bell}}} = \mathrm{i}\tan\left(k\ell + \tfrac{\mathrm{i}}{2}\ln|\mathcal{R}|\right),
  \label{eq:impedance}
\end{equation}
conjugated from the $\mathrm{e}^{-\mathrm{j}\omega t}$ convention of the source. It reduces to $0.25(kr_\text{bell})^2 + \mathrm{i}\,0.6133\,kr_\text{bell}$ as $kr_\text{bell}\to0$. The flanged piston correction in an infinite baffle~\cite{MorseIngard1968} is also available, implemented as a convergent power series in the Struve function $H_1$.

Visco-thermal wall loss is modeled by correcting the lossless real wave number $k_0 = \omega/c$ by the Kirchhoff formula~\cite{Keefe1984} with a real dispersion and an imaginary dissipation term, extended with two empirical parameters ($K_{\mathrm{mat}}$ and $R_a$) that can be used for fine tuning,

\begin{equation}
  k(\omega) = k_0 + k_{\rm disp}(\omega) -\mathrm{i} k_{\rm diss}(\omega),
  \label{eq:kcomplex}
\end{equation}
with
\begin{align}
  k_{\rm disp}(\omega) &= \alpha_0(\omega)\left[1 + \Theta\right] \\
  k_{\rm diss}(\omega) &= \alpha_0(\omega)\left[R(\omega) + \Theta\right],
\end{align}  
where
\begin{align}
  \alpha_0(\omega) &= \frac{K_{\mathrm{mat}}}{rc}\sqrt{\frac{\omega\nu}{2}} \label{eq:kirchhoff} \\
  \Theta &= \frac{\gamma-1}{\sqrt{\text{Pr}}}
\end{align}
Here $r$ is the local tube radius, $\nu = \mu/\rho$ is the kinematic viscosity, $\gamma = 1.4$, $\text{Pr} = 0.71$. The dynamic viscosity $\mu$, air density $\rho$ and speed of sound $c$ are computed from the ambient temperature $T$ by the standard fits, accurate to within 1\% over the range $0$--$40\,^\circ$C and reproduced in the supplementary material. 

$R(\omega)$ is a frequency-dependent correction to account for the surface-roughness of the material the bore is made of, modeled as
\begin{equation}
  R(\omega) = 1 + \frac{R_a}{\sqrt{\delta_v^2 + R_a^2}},
  \label{eq:rough_sat}
\end{equation}
where $\delta_v(\omega) = \sqrt{\frac{2\nu}{\omega}}$ is the viscous boundary layer thickness and $R_a$ is the RMS surface roughness of the bore wall. This term is not derived from first principles. It is chosen as an interpolation between the ideal smooth ($R_a \ll \delta_v$) and fully-rough limits ($R_a \gg \delta_v$), preserving in both cases the correct physical limit $\alpha \propto \sqrt{\omega}$.

Note that $R(\omega)$ appears in the dissipative part of \cref{eq:kcomplex} but not in the dispersive one, so the real and imaginary corrections are not equal in magnitude and the familiar $k = k_0 + \alpha(1-\mathrm{i})$ does not hold once the wall is rough. The reason is that roughness adds drag in the viscous sublayer, which damps the wave, but it does not slow the plane wave by the same factor. The two parameters are calibrated against the measured quality factors of the instruments of \cref{sec:validation}.

The materials offered by the app differ in their surface roughness $R_a$ and in $K_{\mathrm{mat}}$, a residual multiplier covering wall mechanical compliance and any loss mechanism not captured by surface roughness; the values are listed in the supplementary material. The $R_a$ values are order-of-magnitude estimates for each machining class and may be treated as a single tunable wall-loss parameter. $K_{\mathrm{mat}}$ is set to 4 to model soft tissue in case the app is used to demonstrate vocal tract acoustics, is unity for the smooth rigid materials, and is $0.80$ for wood, the only material acoustically calibrated (\cref{sec:validation}). \citet{Boutin2017} report that acoustic dissipation in wooden pipes is measurably material-dependent, consistent with the loss being carried by more than the Kirchhoff boundary layer alone.

\begin{figure}[htb]
    \centering
    \includegraphics[width=\columnwidth]{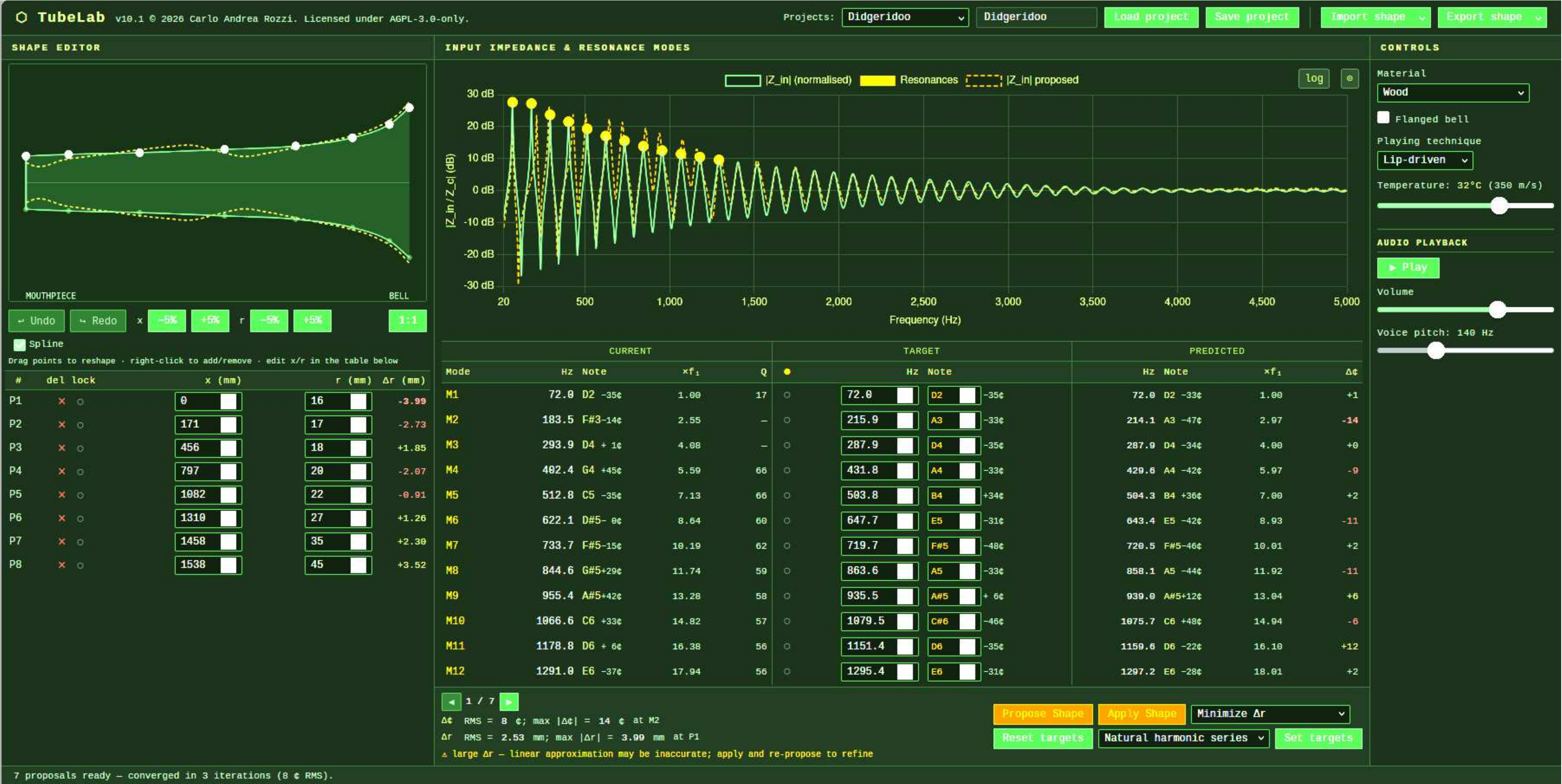}
    \caption{Screenshot of \tl{}. On the left a didgeridoo shape is defined via a few control points. On the right the magnitude of the input impedance is calculated. The user has requested to reshape the bore to fit natural harmonics. The proposed new shape and impedance are overlaid in orange over the original ones. The user can navigate seven different options. Note that the requested target exceeds the linear modeling limit and a warning is issued in the bottom line to iterate the procedure.}
    \label{fig:tl_screenshot}
\end{figure}

After $|Z_\text{in}|$ is calculated its peaks or troughs are detected by a simple three-point inequality test. The positions of the extrema are then refined by golden-section search on the continuous $|Z_\text{in}|$, so they do not depend on the frequency bin size. Quality factors $Q$ are estimated from the $-3$\,dB bandwidth of the impedance magnitude, located on the same continuous curve. The half-power frequencies are bracketed by a scan outward from the peak bin, and each crossing is then refined by bisection on that same continuous curve. The first 12 resonance frequencies are reported and are selectable for tuning (see \cref{sec:inverse}). The user is warned if any mode falls beyond the bell cut-off frequency (see \cref{fig:tl_screenshot}).

Mode shapes are overlaid on the bore geometry outline when the user hovers on a resonance in the impedance plot. Each shape is computed by a forward TMM pass from mouth to bell at the resonance frequency $f_n$, seeded with the mouth boundary condition ($p=1,\,U=0$ for brass/reed excitation; $p=0,\,U=1$ for edge-blown), with $|p(x)|$ normalised to unity. This is a very important piece of information for makers, who are generally aware of the effect on individual modes of widening or narrowing the bore close to nodes or anti-nodes. The missing piece of information, i.e., how much pointwise modifications affect the {\em other} modes is obtained by addressing the inverse problem.

\section{Bore inversion}
\label{sec:inverse}

\subsection{Shape sensitivity}
A common inverse problem formulation reads: given the current bore profile $\mathbf{r}^0 = \{r^0_i\}$ with $i\in[1,N]$ and its resonance frequencies $\{f^0_n\}$ with $n\in[1,m]$, and given at most $m$ target frequencies $\mathbf{f}^* = \{f^*_n\}$, find a profile correction $\Delta \mathbf{r}$ such that the shifted resonances satisfy $\mathbf{f}(\mathbf{r}^0 + \Delta \mathbf{r}) \approx \mathbf{f}^*$. Defining the target frequency shift $\Delta\mathbf{f}^* = \mathbf{f}^* - \mathbf{f}^0$ and linearizing for small deviations, the problem becomes a linear least squares minimization
\begin{equation}
  \min_{\Delta \mathbf{r}} \left\| J\Delta \mathbf{r} - \Delta\mathbf{f}^* \right\|^2,
  \label{eq:leastsq}
\end{equation}
where $J$ is the $m\times N$ sensitivity Jacobian with elements $J_{ni} = \partial f_n/\partial r_i$.

$J_{ni}$ is computed by the phase-perturbation method differentiating the resonance condition $\operatorname{Im}(Z(f;\mathbf{r})) = 0$ at $f = f_n^0$ with respect to the profile perturbation $\delta r_i$ to get
\begin{equation}
  J_{ni} = \frac{d f_n}{d r_i}
  = -\left.\frac{\partial_{r_i} \operatorname{Im}Z}
           {\partial_{f_n} \operatorname{Im}Z}\right|_{f_n^0}.
  \label{eq:Jift}
\end{equation}
The denominator is precomputed once per mode via a centered finite difference of the
base profile ($\pm 0.5$\,Hz). Each perturbation evaluation then requires only a single-frequency TMM computation at $f_n^0$, rather than a full impedance sweep. The numerator is computed via the adjoint-state method~\cite{Plessix2006}: one forward $O(N)$ TMM pass that caches the per-segment state, followed by one backward co-state $O(N)$ pass that accumulates $\partial\operatorname{Im}Z/\partial r_i$ for all $N$ profile points simultaneously.

\subsection{Constraints}
We improved \cref{eq:leastsq} in four ways. 

First, since pitch perception is roughly logarithmic it makes more sense to minimize the error in log frequency (cents) instead of linear frequency (Hz). This is obtained by rescaling the Jacobian by the base resonances
\begin{equation}
  \widetilde{J}_{ni} = \frac{J_{ni}}{f_n^0},
  \label{eq:Jlog}
\end{equation}
and expressing the target vector in nepers
\begin{equation}
  \tilde{f}_n = \ln\left(\frac{f^*_n}{f^0_n}\right).
  \label{eq:dflog}
\end{equation}

Second, since~\eqref{eq:leastsq} is strongly under-determined ($m \leq 12$ equations, $N = 200$ unknowns), a Tikhonov regularization term is added to select among infinitely many solutions
\begin{equation}
  \min_{\Delta\mathbf{r}}
  \left[
  \left\|
    \widetilde{J}\,\Delta\mathbf{r} - \tilde{\mathbf{f}}
  \right\|^2
    + \lambda\,\Delta\mathbf{r}^T \Gamma\,\Delta\mathbf{r}
  \right],
  \label{eq:tikhonov}
\end{equation}
where $\lambda > 0$ weights the penalty against the data term and $\Gamma$ is a symmetric positive semi-definite matrix that decides which corrections are to be considered expensive. The resulting normal equations
\begin{equation}
  (\widetilde{J}^T\widetilde{J} + \lambda \Gamma)\,\Delta\mathbf{r} = \widetilde{J}^T\tilde{\mathbf{f}}
\end{equation}
can be solved by Gaussian elimination with partial pivoting. Two choices of $\Gamma$ are implemented, expressing different notions of a costly bore change: $\Gamma = I$ penalizes the size of the correction, minimizing $\sum_i \Delta r_i^2$, while the tridiagonal $\Gamma = L^T L$ built from the first-difference operator $L$ penalizes its unevenness, minimizing $\sum_i (\Delta r_{i+1} - \Delta r_i)^2$.

The Tikhonov parameter $\lambda$ controls the balance between frequency accuracy and maximum shape variation. A choice among 7 solutions, generated from the same Jacobian by sweeping logarithmically $\lambda\in[10^{-2},10^3]\|\widetilde{J}\|^2$, is offered to the users. Larger $\lambda$ produces smoother, smaller-magnitude corrections that may not reach the target frequencies; smaller $\lambda$ reaches the targets more closely but may produce aggressive bore changes. The user can navigate between the variants to accept whichever offers the best trade-off. Incidentally, working with frequency-scaled Jacobian improves $\lambda$ auto-range accuracy: if $\|\widetilde{J}\|^2$, were in Hz/mm instead of 1/mm, it would be dominated by high-frequency modes, and $\lambda$ would be forced to larger values, over-regularizing contributions from low-frequency modes.

Third, makers might be bound to geometry constraints. For example they might \emph{not} want to modify some parts of their initial design ($\Delta r_i = 0$ for some $i$), or might be allowed to widen, but not narrow the bore via gouging ($\Delta r_i \ge 0\, \forall i$). The first constraint is imposed by simply removing the corresponding points from the $\Delta \mathbf{r}$ vector. The second is imposed iteratively by identifying negative $\Delta \mathbf{r}$ components in a solution, excluding their columns from the solve and re-solving with diagonal regularization.

Last, during the optimization a maker might be already satisfied with the tuning of \emph{some} modes, but not with others. An equality boundary must be made available to lock those modes, restricting the solution to the subspace of the {\em free} modes. Hard constraints are enforced via the Karush--Kuhn--Tucker (KKT) saddle-point system \cite{Ruszczynski2006} by separating $J$ into free ($J_F$) and locked ($J_L$) block matrices and solving
\begin{equation}
  \begin{pmatrix}
    A & B^T \\
    B & 0
  \end{pmatrix}
  \begin{pmatrix} \Delta\mathbf{r} \\ \bm{\xi} \end{pmatrix}
  =
  \begin{pmatrix} \widetilde{J}_F^T\,\tilde{\mathbf{f}}_F \\ \mathbf{0} \end{pmatrix},
  \label{eq:kkt}
\end{equation}
where $A = \widetilde{J}_F^T\widetilde{J}_F + \lambda \Gamma$, $B = \widetilde{J}_L$, and $\bm{\xi}$ are Lagrange multipliers. The KKT system is linear, so the locked-mode constraint is exact to first order. Nonlinear residuals are reduced by up to four iterations of a minimum-norm correction step; the tolerance is 0.5\,Hz. If residual drift still exceeds this threshold after four iterations --- which occurs only for large bore corrections where the linear model is already a poor approximation --- the app displays an explicit warning prompting the user to apply the current correction and re-propose, effectively replacing a formal convergence criterion with an interactive refinement loop.

The approach closest to ours is that of \citet{Ernoult2020}, who also use the adjoint-state method within a TMM recursion to compute resonance-frequency sensitivities and enforce geometric constraints on the bore. Several aspects of the present formulation differ in ways relevant to the interactive, maker-facing use case: 1) our target specification is a discrete set of resonance frequencies, not a full impedance cost function. This allows users to express musically meaningful objectives (note names, interval ratios, tuning systems) on individual modes without specifying the entire impedance curve. This formulation incidentally simplifies the inverse solution to one forward pass, one adjoint Jacobian evaluation, $O(mN)$ for $m \leq 12$ modes over $N = 200$ profile points, and one tiny KKT solve; 2) the parameter design space is different: \tl{} operates in the 200-dimensional profile-radius space, which enlarges the solution null-space to accommodate hard frequency-lock equality constraints without ill-conditioning. In fact selected modes can be enforced as exact hard equality constraints via the KKT saddle-point formulation. Regardless of the Tikhonov parameter $\lambda$, the locked-mode residual is driven exactly to zero by the constraint structure rather than approximately by a penalty weight; 3) instead of a single automated optimization run, the user can navigate a family of seven differently regularized solutions, choosing whichever bore correction best balances spectral accuracy against shape magnitude or smoothness for the specific instrument at hand; 4) the entire workflow runs in a browser with no installation required, making it accessible to instrument makers without specialist numerical training. 

A few more handlers further facilitate usage: resonance frequencies, in app, are also represented in music notation (note + cents with respect to equal temperament), and common sets of musical target are hard-coded, including popular tuning systems such as the natural harmonic series, Pythagorean fifths, just fifths and thirds, Bohlen-Pierce tritave, neutral intervals \emph{maqam}. Moreover a few common exploration options (rescale radii, rescale length, shape modification undo-redo, etc.) are quickly available and require no file manipulation. The app also supports loading and saving profiles as CSV and AutoCAD DXF formats, to enable the use of numerical control (NC) machines (see \cref{fig:tl_screenshot}).

\section{Validation}
\label{sec:validation}

\tl{} includes five preset geometries, mostly meant for educational purposes, covering simplified shapes for common instruments, and seven vocal-tract profiles. The latter were actually obtained from the inverse solution provided by the app itself by imposing tabulated formants for the Italian vowels \cite{Ferrero1986}. We used these presets for the purely numerical benchmark.

For the experimental validation we employed a set of commercial didgeridoos with varied sizes and non-trivial shapes, all fabricated by A.\,F. using a NC machine. Each nominal CAD profile defines a class (A--F). Each class comprises the individual specimens milled from that profile. Six classes were available, realized as sixteen specimens (respectively 1, 2, 4, 6, 1, 2 specimens per class). Class A has the most extreme conical profile, meant to tune the overtones close to harmonic ratios ($f_2/f_1 = 2.00$, $f_3/f_1 = 2.95$), while across the six classes $f_2/f_1$ spans 1.96--2.56. Classes~B and C are two differently tuned versions of a design meant to enhance vocal-tract and formant interaction; Class~D was designed with increased back pressure to facilitate learning; Class~E was designed for a brighter timbre, emphasizing resonances close to the third and the octave. Class F was optimized to balance timbre richness and playability.

The bores themselves are commercial designs and are not reproduced here. However the classes are geometrically well separated (see the scale-free descriptors in \cref{tab:instruments}). The single exception is the pair B/C, which share the same flare profile at two different scales; they are nevertheless distinct classes because they are separately tuned and separately machined.

\begin{table}[ht]
    \centering
  \caption{Scale-free characterization of the six nominal bores. $\theta$ is the area-weighted mean wall half-angle, $\theta = \int|\!\arctan(\mathrm{d}r/\mathrm{d}x)|\,S\,\mathrm{d}x \,/\, \int S\,\mathrm{d}x$; $r_\text{bell}/r_\text{mouth}$ is the overall flare; $f_2/f_1$ is measured. $\theta$ is the covariate of the $\kappa\theta^2$ fit of \cref{sec:numbench}. The set spans a factor 5.5 in $\theta$, and class~A is the most flared of the eleven geometries tested against FEM, the five built-in presets included. Classes B and C are one design at two lengths: their flare ratios agree to 0.2\,\% while their $\theta$ differ, $\theta$ being sensitive to the aspect ratio and the flare ratio not.}
  \label{tab:instruments}
  \begin{tabular}{lccc}
  \toprule
  Class & $\theta$ & $r_\text{bell}/r_\text{mouth}$ & $f_2/f_1$ \\
        & (deg)     &                                 &           \\
  \midrule
  A  & 2.59 & 4.05 & 2.00 \\
  B  & 0.79 & 2.11 & 2.56 \\
  C  & 0.95 & 2.10 & 2.54 \\
  D  & 0.47 & 2.33 & 1.96 \\
  E  & 0.66 & 2.25 & 2.51 \\
  F  & 1.69 & 2.78 & 2.03 \\
  \bottomrule
  \end{tabular}
\end{table}

\subsection{Numerical benchmarks}
\label{sec:numbench}

The accuracy of the app direct problem solutions was first assessed against FEM calculations of the input impedance performed in a 2D axisymmetric arrangement within the pressure acoustics module of COMSOL Multiphysics\textregistered\ \cite{Comsol64}. The FEM model was forced to impose the same bell radiation impedance (\cref{eq:radimpedance_unflanged}), the same rigid mouth and the same wall-loss law as the TMM. The agreement therefore bounds the interior solver.

Grid adequacy was verified by a convergence study comparing TMM resonances against an $N=1200$ reference: all six didgeridoo profiles reach $\Delta f_\text{RMS} < 1$~cent by $N = 100$ grid points, and at the production setting of $N = 200$ the residual is at most 0.27~cent RMS on any of them (0.56~cent on the worst single mode). The direct calculations take a mean wall-clock time at $N = 200$ in the range 30--60\,ms on modern PC hardware, which places the full interactive update within real-time response.

The comparison covers the first twelve modes of both the five built-in demonstration instrument presets and the six commercial nominal didgeridoo bores, for a total of 132 mode pairs, tested in both the lossless and lossy conditions. 

\begin{figure}[htbp]
    \centering
    \includegraphics[width=\columnwidth]{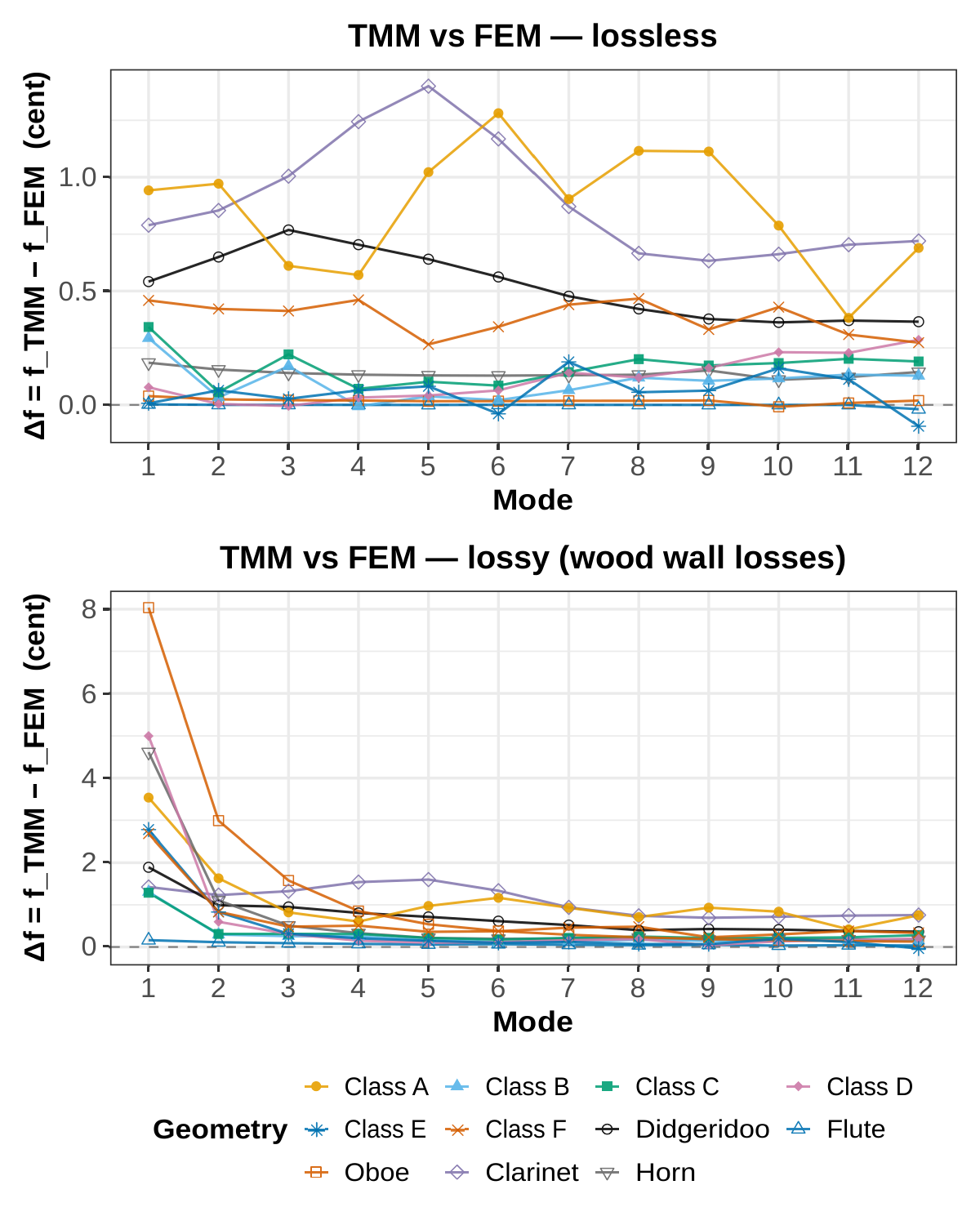}
    \caption{Comparison of the resonance frequencies predicted by \tl{} TMM and Comsol FEM at the same physics approximation level on the same set of geometries. The deviation is $\Delta f = 1200\log_2(f_\text{TMM}/f_\text{FEM})$, so a \emph{positive} value means the TMM places the resonance sharper than the FEM. Note the different vertical scales: the lossless panel (top) spans about one cent and is flat in mode index, whereas the lossy panel (bottom) carries an excess at the fundamental that decays monotonically and is absent above. The wall-loss term is the only physics that differs between the two.}
    \label{fig:tmm_comsol_delta}
\end{figure}

In the lossless case (\cref{fig:tmm_comsol_delta}, top) the RMS deviation over the 132 mode pairs is 0.45 cent, and it is flat in mode index. The per-geometry bias is just $-0.001$ cent for the flute, the one exactly cylindrical bore of the set, and it grows with the area-weighted mean wall half-angle $\theta$ of the bore (\cref{tab:instruments}) as $\Delta f = \kappa\theta^2$ with $\kappa = 442$ cent/rad$^2$ ($R^2 = 0.95$), fitted through the origin over all eleven geometries; $\theta$ for each of them is tabulated in the supplementary material. The interpretation is the plane-wave assumption made in TMM: in a flaring bore the wavefront is curved, so the acoustic path exceeds the axial length that the one-dimensional TMM integrates along, and the FEM, which resolves the two-dimensional field, places the resonance correspondingly lower.

In the lossy case (\cref{fig:tmm_comsol_delta}, bottom) the RMS is 1.21 cent, and it is not uniform: 3.7 cent at the fundamental, falling steadily to 0.38 cent at the twelfth mode (0.63 cent over modes 2--12), with the same sign throughout, the TMM always placing the resonance sharper. Since the wall-loss term is the only physics that differs between the two cases, the excess is attributable to it, and specifically to the fact that the two codes represent it differently: \tl{} carries the loss in the complex wavenumber of a one-dimensional plane wave (\cref{eq:kcomplex}), while the FEM imposes it as a wall impedance $Z_n = \rho c / [r(k_\text{diss} + \mathrm{i}\,k_\text{disp})]$ on the boundary of a two-dimensional axisymmetric domain, where the resulting field is not exactly a plane wave. The two formulations agree to first order in $\alpha/k_0$. Consistent with that reading, the excess is largest where damping is strongest and at the lowest modes. 

$Q$ factors agree to a median of 1.59\% in the lossy case. Agreement across the full impedance curve was separately verified for the six commercial didgeridoos: the $|Z|/Z_c$ magnitude (dB) over 20--5000\,Hz matches COMSOL with a pooled RMS deviation of 0.19\,dB and a Pearson correlation of 1.000; over the musically relevant 20--1500\,Hz band the RMS deviation reduces to 0.15\,dB, confirming that agreement extends continuously across the full audible spectrum, not only at individual resonance peaks. 

Vocal tract deviations are larger: 18.6 cent for the resonance frequencies (16.4 cent lossless); these figures are still in good agreement given the more challenging geometry, but reveal that short tubes with strongly varying profiles are more demanding for a simple plane-wave approach. Nevertheless, the generated shapes resemble known ones from radiologic measurements and, most importantly, they provide realistic approximation to the vowels for a native Italian speaker when their sound is simulated in app (see \cref{sec:synthesis}).

A second, independent cross-check was made against OpenWind~\cite{Chabassier2020OpenWinD}, a finite-element frequency-domain solver. While in the COMSOL comparison we forced the same radiation and loss models, to validate the interior solver alone, OpenWind brings its own models and spherical-wave conical elements. Resonances are compared on the $\operatorname{Im}(Z) = 0$ definition for both codes, which is OpenWind's own and differs from the $|Z|$ extremum used elsewhere in this work; the offsets quoted below are relative to that common definition.

Over the six commercial bores the two agree to within a smooth monotone offset rising from 1.4 cent at the fundamental to 4.6 cent at the twelfth mode, with no structure in between. Switching OpenWind's physics one option at a time we find that replacing the exact Bessel losses by the first-order form \tl{} uses moves every mode by less than $0.01$ cent, spherical-wave elements by $0.6$ cent flat in mode index, and the causal radiation Pad\'e by at most $0.9$ cent at the twelfth mode. The near-exactness of the first shows that the higher-order thermoviscous terms, second order in the ratio of the viscous boundary-layer thickness to the bore radius, are irrelevant at these radii. 

\subsection{Experimental validation}

For real-world didgeridoos we measured resonance frequencies following a protocol that also makers could replicate in their workshops: the room temperature was noted (24$^\circ\pm 0.5^\circ$ C for the whole present data set); the tube was rested on soft foam; the mouth was covered with rigid tape; a unidirectional mic was placed at about 5 cm from the bell in an enclosure covered with sound absorbing material; the mouth was excited tapping lightly on the covering tape; at least ten taps per instrument were recorded at 48 kHz and saved as wav files. The taps of a given specimen are averaged in the spectral domain and each resonance is fitted with a Lorentzian, so every specimen yields its own resonance list. The analysis window is matched to each mode depending on amplitude decay time of the mode being fitted to optimize the SNR of each tap. Since $Q$ is unknown before fitting, the window is selected iteratively from the fitted linewidth. 

The two free parameters of the loss model, $K_{\mathrm{mat}}$ and $R_a$, are fitted from these same tap responses. The measured $Q$ is estimated from sound emission and it includes several loss channels: $Q_\text{tot}^{-1} = Q_\text{wall}^{-1} + Q_\text{rad}^{-1} + Q_\text{other}^{-1}$. Only the first depends on the parameters being fitted, so the radiation channel is removed from both sides using the model's own lossless run. Every channel left in $Q_\text{other}$ --- the tape seal, the foam support, transmission through the wall --- and the finite-window broadening of the Lorentzian fit all \emph{depress} the measured $Q$, so these are lower bounds on the over-damping, not estimates of it.

A two-parameter grid search against the radiation-corrected $Q$ returns $K_\mathrm{mat} = 0.80$ ($1\sigma$ interval $0.75$--$0.80$) and $R_a = 60\,\mu$m; at that $K_\mathrm{mat}$, setting $R_a = 0$ is 38\,\% worse, so the roughness term itself is required by the damping data and what the fit determines is its scale. The damping data does not, however, constrain where the roughness factor sits: the fit objective changes by 0.4\,\% whether $R(\omega)$ is confined to the dissipative part of \cref{eq:kcomplex} or carried into both. That placement rests on the physical argument of \cref{sec:physics} alone, and it moves the predicted fundamental by 2.7 cent on average over the six bores, about half of the 4.5 cent that the calibration of $K_\mathrm{mat}$ itself is worth.

\begin{figure}[htbp]
    \centering
    \includegraphics[width=\columnwidth]{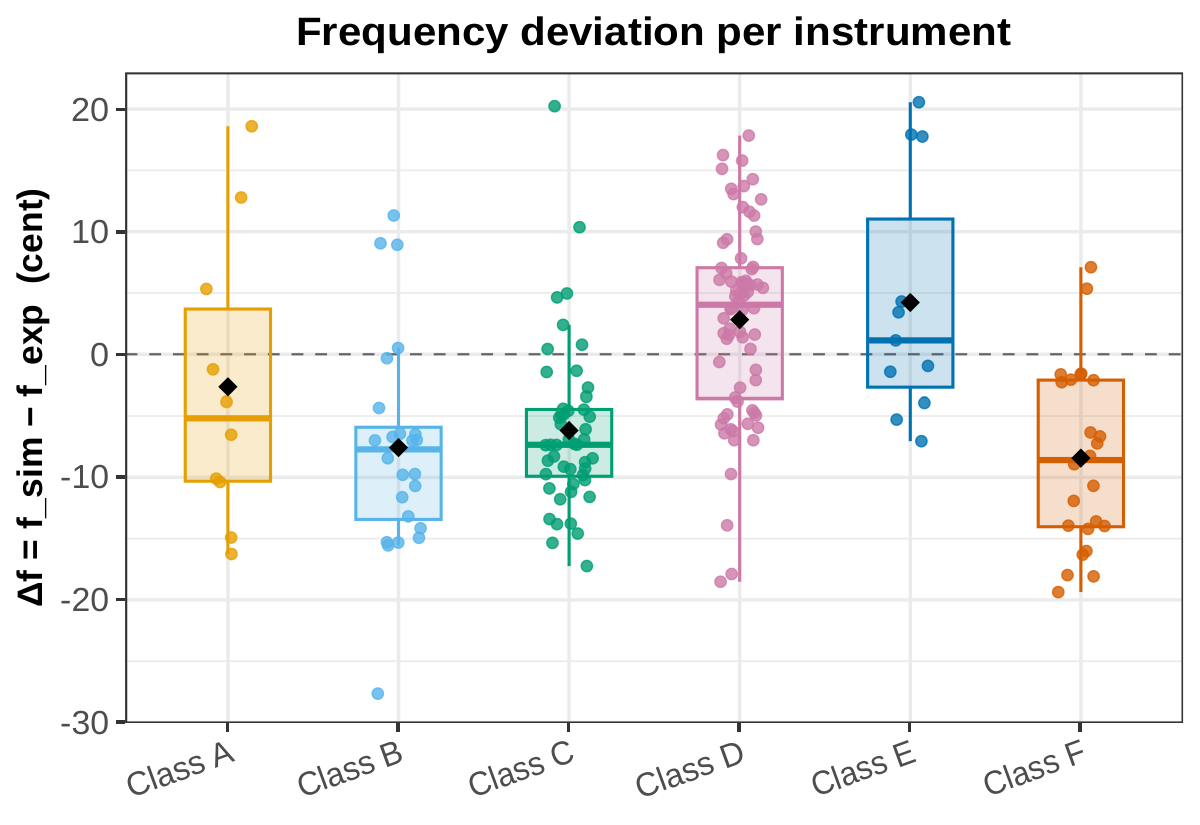}
    \caption{Per-class and overall deviations of simulated from measured resonance frequencies. Diamonds mark the per-class mean. The data set includes the first twelve modes of sixteen specimens spanning six classes of commercial didgeridoo fabricated with a NC machine; each specimen contributes its own point, all specimens of a class being compared against the single nominal CAD bore of that class. The mean deviation is $-2.4$ cent and the pooled RMS deviation is 9.6 cent over 189 matched pairs.}
    \label{fig:resonance_comparison}
\end{figure}

The results are summarized in \cref{tab:freq_comparison} and \cref{fig:resonance_comparison}. The mean deviation of the predicted from the measured frequencies is $-2.4$ cent and the pooled RMS deviation is 9.6 cent over 189 matched pairs. Since no frequency data entered the calibration just described, the frequency agreement reported here is a prediction of that calibration rather than a fit to the quantity being reported.

\begin{table}[ht]
  \centering
  \caption{Per-class frequency accuracy, $\Delta f = f_\text{sim} - f_\text{exp}$. Each of the $n_\text{spec}$ specimens of a class is compared against the same nominal CAD bore; RMS is computed over all matched pairs of that class and the overall row pools all pairs.}
  \label{tab:freq_comparison}
  \begin{tabular}{lccccc}
  \toprule
  Class & $n_\text{spec}$ & n. pairs & $\overline{\Delta f}$ & $|\Delta f|_\text{rms}$ & $|\Delta f|_\text{max}$ \\
  & & & (cent) & (cent) & (cent) \\
  \midrule
  A  & 1 &  10 & $ -2.7$ & 11.4 & 18.6 \\
  B  & 2 &  24 & $ -7.6$ & 11.5 & 27.6 \\
  C  & 4 &  48 & $ -6.2$ &  9.1 & 20.2 \\
  D  & 6 &  72 & $ +2.8$ &  8.3 & 18.5 \\
  E  & 1 &  11 & $ +4.2$ & 10.4 & 20.6 \\
  F  & 2 &  24 & $ -8.4$ & 11.1 & 19.4 \\
  \midrule
  Overall & 16 & 189 & $ -2.4$ &  9.6 & 27.6 \\
  \bottomrule
  \end{tabular}
\end{table}

\begin{figure}[htbp]
    \centering
    \includegraphics[width=\columnwidth]{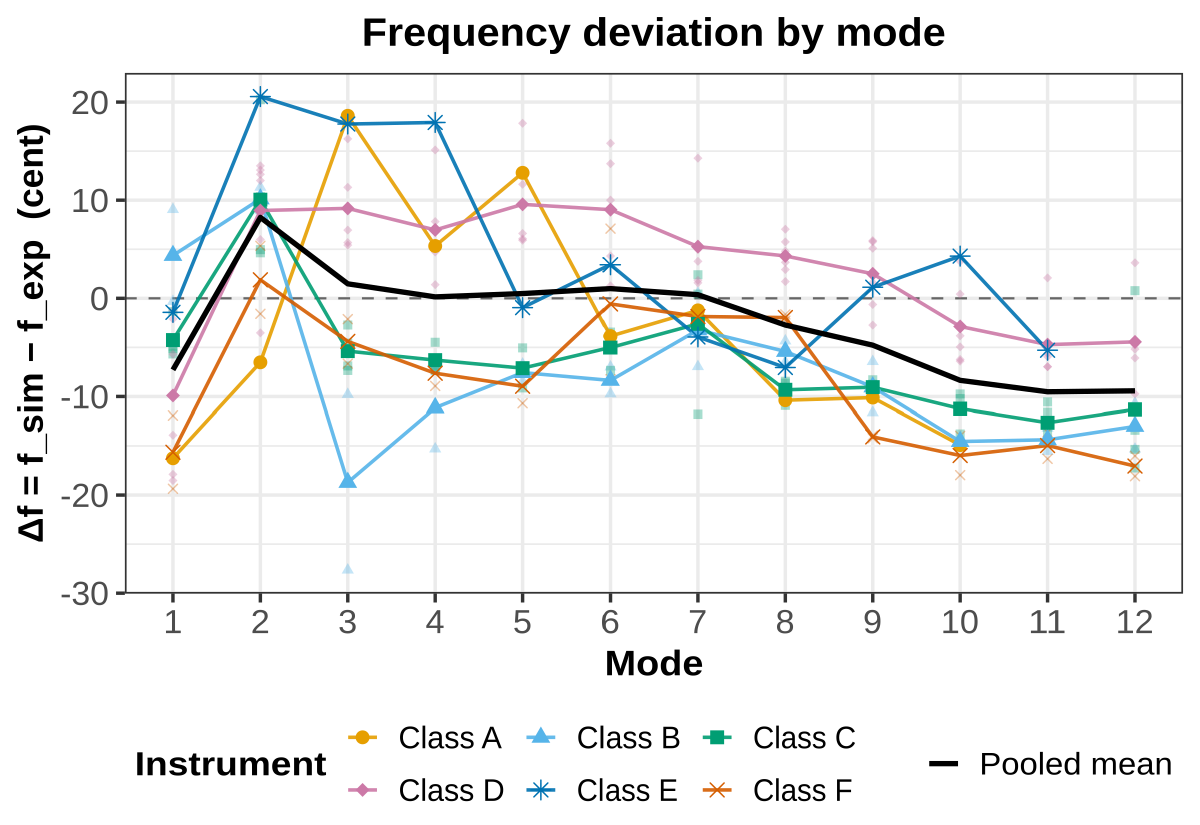}
    \caption{Frequency deviation against mode index. Faint markers are the individual specimens; the coloured lines join each class's mean at each mode; the heavy black line is the mean pooled over all classes. The two structures discussed in the text are separated here: the colored lines fan out over modes 3--5, where the between-class spread reaches 10.9 cent about a pooled mean of only $+0.7$ cent, and converge below zero over modes 9--12.}
    \label{fig:resonance_by_mode}
\end{figure}

To deepen the analysis we examined the per-mode deviations. \cref{fig:resonance_by_mode} shows three different trends depending on the mode order. The high modes 9--12 carry a common offset shared by the classes (mean $-7.9$, between-class spread $6.3$ cent), whereas the middle of the spectrum carries almost no common offset and is dominated by structure specific to each bore (mean $+0.7$~cent, spread $10.9$~cent). 

Modes 1 and 2 are peculiar. They are biased in opposite directions in all six classes ($-7.3$ and $+8.2$ cent respectively). Stated in the form that matters to a player, the model over-predicts the overblow interval $f_2/f_1$ by 15.5 cent. It is not an artifact of this implementation, though. OpenWind reproduces it at $-5.3$ and $+10.0$ cent, agreeing with \tl{} elsewhere to within the smooth offset quoted above. It is not the measurement either: injecting modes of known frequency into the extraction pipeline, at the measured signal-to-noise ratio and with the recordings' own room noise, recovers them to $-0.11$ cent at mode 1 and $+0.04$ at mode 2. We report it as a limitation shared by both one-dimensional models.

\subsection{Fabrication reproducibility}
\label{sec:withinclass}

Four of the six classes are represented by two or more specimens milled from the same CAD file. How much those copies disagree with each other is a property of the production process alone: it involves no acoustic model, and it can be measured because each specimen was tapped and analyzed independently.

\begin{figure}[htbp]
    \centering
    \includegraphics[width=\columnwidth]{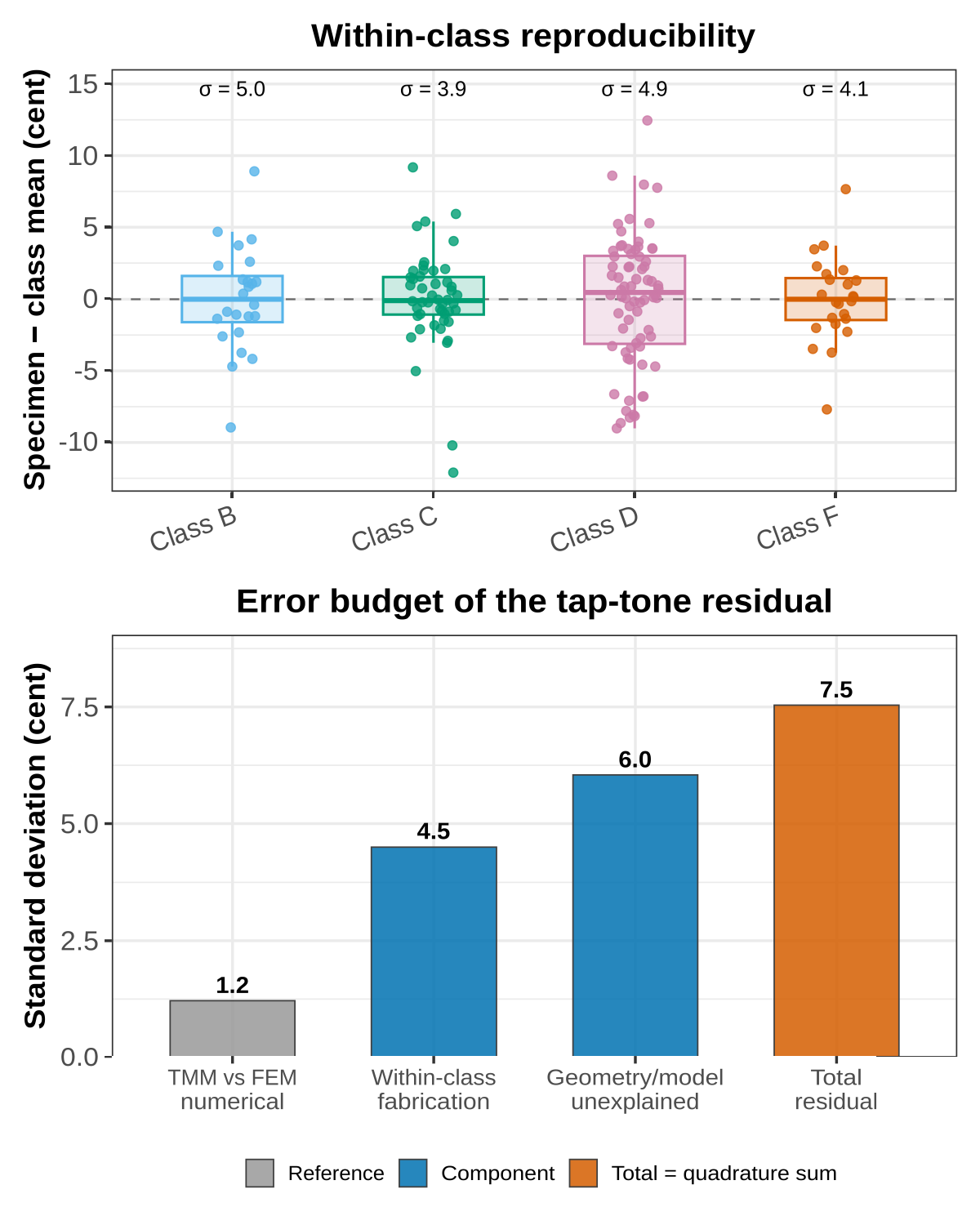}
    \caption{Top: deviation of each specimen from the mean of its own class and mode, for the four classes having more than one specimen; $\sigma$ is the per-class pooled within-class scatter, whose pooled value over all four classes is $\sigma_\text{fab} = 4.5$ cent. Bottom: decomposition of the simulation-vs-experiment residual. Fabrication scatter alone accounts for 36\,\% of the residual variance. The gray bar --- the RMS disagreement between the TMM and FEM on identical geometry --- is drawn for scale only and is not a term of that sum.}
    \label{fig:within_class}
\end{figure}

For every (class, mode) cell we computed the standard deviation of the measured resonance frequency across the class's specimens. Pooling the 48 available cells gives a within-class reproducibility of $\sigma_\text{fab} = 4.5$ cent, with per-class values of 3.9, 4.1, 4.9 and 5.0 cent (\cref{fig:within_class}, top). The spread is flat in cents across mode index --- it ranges between 1.4 and 6.9 cent per mode with no trend --- so fabrication perturbs each resonance by an independent relative amount.

The data are nested --- six designs, several copies of most of them, up to twelve resonances per copy --- so a mixed-effects model fitted over all 189 pairs can attribute the deviation of simulation from experiment to the level at which it arises (class, specimen, mode):
\begin{equation}
  \Delta f_{ijn} = \beta_0 + \beta_1 \log (f_{ijn}/f_{i0})
                 + u_i + v_{ij} + \varepsilon_{ijn},
  \label{eq:mixed}
\end{equation}
where $i$ indexes the class, $j$ the specimen within that class and $n$ the mode. The covariate is referred to $f_{i0}$, the simulated fundamental of class $i$, so that it measures how far up its own instrument's spectrum a resonance sits rather than that instrument's absolute pitch; the three random terms are independent and normal with standard deviations $\sigma_\text{class}$, $\sigma_\text{spec}$ and $\sigma_\text{resid}$ respectively. The specimen-level intercept $v_{ij}$ is small and not distinguishable from zero ($\sigma_\text{spec} = 1.9$ cent, with a restricted maximum likelihood profile interval running from 0 to 4.3 cent): to the accuracy of this data set, specimens of a class do not differ by a constant pitch offset and none of them is systematically sharp or flat as a whole. The mode-level residual (the standard deviation of $\varepsilon_{ijn}$, that is the scatter of one individual resonance about the prediction made for its own specimen) carries most of the variance: $\sigma_\text{resid} = 7.5$ cent with a profile interval of 6.8 to 8.4 cent. The class-level intercept $u_i$, by contrast, is clearly resolved at $\sigma_\text{class} = 4.8$ cent with a profile interval of 2.0 to 9.5 cent: the six designs do differ from one another by more than their specimens differ among themselves. Fabrication scatter therefore lives entirely at the level of the individual resonance, which is what one expects from local departures of the milled bore from the CAD profile rather than from a global scaling error of the machine.

The mode-level residual then can be decomposed in quadrature into two terms: the within-class fabrication scatter $\sigma_\text{fab} = 4.5$ cent measured above, and a remainder $\sigma_\text{geo} = \sqrt{\sigma_\text{resid}^2 - \sigma_\text{fab}^2} = 6.0$ cent which carries geometry error and model deficiency together (\cref{fig:within_class}, bottom). The two latter components cannot be further separated without a direct measurement of the bore of the finished instrument. 

\subsection{Inverse validation}

The inversion capability of \tl{} can be exploited to estimate the modifications predicted by the model to force the bores to perfectly match their \emph{measured} frequency targets starting from simulated ones, and comparing them to intrinsic geometry uncertainty. The latter is affected by many factors: besides the $\pm 0.2$\,mm CAD-to-NC tolerance quoted for the machine that covers the cutting step alone, also tool alignment, internal surface finishing, coating or varnishing, internal stress and moisture-driven movement contribute to alter the finished bore. 

The bore corrections were obtained by performing a 10 point $\lambda$ log sweep and taking the solution whose residual is closest to 3 cent, with no further refinements. Pooled over the six classes, matching the measured frequencies to 3.1 cent RMS requires bore corrections of 0.36 mm RMS.

\begin{table}[ht]
  \centering
  \caption{Inversion results per class taking as target the experimental frequencies of the first 12 modes. $|\Delta r|_\text{fab}$ is the rms bore difference between two physical copies of the same design, obtained by passing that class's measured fabrication scatter through the same Jacobian at the same $\lambda$; it is defined only for the four classes having more than one specimen.}
  \label{tab:inversion}
  \begin{tabular}{l|cc|cc}
  \toprule
  Class & $|\Delta f|$ & $|\Delta r|_\text{max}$ & $|\Delta r|_\text{rms}$ & $|\Delta r|_\text{fab}$ \\
  & (cent) & (mm) & (mm) & (mm) \\
  \midrule
  A  & 3.3 & 1.82 & 0.66 & ---  \\
  B  & 2.6 & 0.70 & 0.27 & 0.25 \\
  C  & 3.2 & 0.51 & 0.20 & 0.20 \\
  D  & 2.4 & 0.67 & 0.08 & 0.07 \\
  E  & 3.7 & 0.90 & 0.18 & ---  \\
  F  & 3.3 & 1.03 & 0.42 & 0.57 \\
  \midrule
  B,C,D,F  & 2.9 & 1.03 & 0.27 & 0.23 \\
  All six  & 3.1 & 1.82 & 0.36 & ---  \\
  \bottomrule
  \end{tabular}
\end{table}

In \cref{tab:inversion} $|\Delta r|_\text{rms}$ is the distance from the nominal CAD bore to the \emph{average} physical instrument of its class, since the inversion targets the class-mean measured frequencies. Feeding the same solver, at the same $\lambda$ and through the same Jacobian, with each specimen's deviation from that class mean instead yields $|\Delta r|_\text{fab}$: the distance from the average instrument to an \emph{individual} one. 

The table shows that the geometric adjustment needed to reconcile the simulation with a measured instrument is therefore comparable to the geometric difference the workshop itself introduces between two instruments cut from the same file. As a consistency check, the frequency scatter recovered along this route (5.03, 3.93, 4.85 and 4.12 cent for classes B, C, D and F) reproduces the $\sigma_\text{fab}$ of \cref{sec:withinclass} to within rounding, although the two are computed by entirely independent methods.

\subsection{Sound synthesis}
\label{sec:synthesis}

To separate bore acoustics from player effects, sustained played-tone (drone) recordings were collected for thirteen of the sixteen specimens, and for a few of them on the overblown notes as well, giving 29 takes that span modes 1 to 4. The fundamental pitch was extracted by autocorrelation. The observable is the played note against that specimen's own tap tone at the same mode.

The played note is consistently sharper than the tap, by $+43.8 \pm 14.9$ cent at mode 1 and $+23.2 \pm 9.2$ cent at mode 2. Since a warm bore is very likely present while the instrument is played, for the drone analysis only we adopt an empirical playing temperature of $28\,^\circ$C; the acoustic design path, which rests on tap measurements, keeps the ambient value. Lip compliance and mass, and the coupling of the player's vocal tract to a low input impedance, although relevant when the instruments are played, are out of scope here. 

The goal of this simplified synthesis module is just to give users and makers an immediate auditory impression of the kind of sound the instrument will produce from a given design, rather than a full physics-driven simulation. Therefore we have devised a real-time sound synthesis based on the Web Audio API \cite{WebAudio} just convolving a reasonable harmonic source with the impulse response $h(t)$, which encodes the bore's resonance structure
\begin{equation}
  h(t) = \sum_{n=1}^{N_\text{modes}} A_n \, e^{-t/\tau_n} \cos(2\pi f_n t).
  \label{eq:tubeIR}
\end{equation}
Here $f_n$, $Q_n$ are the resonance frequency and quality factor obtained from the acoustic simulation, $\tau_n = Q_n/(\pi f_n)$ and the mode amplitude is $A_n = w_n \cdot \eta_n^\text{rad}$, where $w_n$ is the normalized peak impedance magnitude, and $\eta_n^\text{rad} = (k_n r_\text{bell})^2 / [1 + (k_n r_\text{bell})^2]$ a rational approximation to the piston radiation resistance, normalized to the first resonance mode.

The harmonic source that models the buzzing lips is a sawtooth oscillator at $f_1$. A few parameters inferred from the recordings add realism. A slight frequency glide from below to $f_1$ over $\sim$15--55\,ms simulates the player's embouchure locking onto the tube resonance. An onset noise burst (band-pass filtered white noise, $Q = 0.8$, decay $\tau = 15$\,ms) models the pre-lock aperiodic phase. A 5.5\,Hz vibrato LFO ramps up over 0.8\,s. The amplitude attack lasts $\approx 13$\,ms. These parameters provide an acceptable feeling for didgeridoo sound.

Agreement with the recorded sound spectra is quantified by computing the Pearson correlation between Gaussian-smoothed ($\sigma = 2$ harmonics) recorded and simulated amplitudes in dB. Pooling the thirteen drone recordings, spanning all six classes, gives a mean envelope correlation of 0.91 (range 0.83--0.98).

Three more excitation models (edge blown, reed and voice) are implemented in app for demonstration purposes of other instruments.

\section{Conclusions}

We have presented \tl{}, a browser-based acoustic simulator and inverse bore-design tool for wind instruments without tone holes and validated it against sound response measurements on a set of didgeridoos built in a controlled way. We highlight three results of methodological interest. 

First, building on the adjoint-state formulation for the TMM, we show that when the design target is a discrete set of resonance frequencies rather than a full impedance cost function, the gradient collapses to a single forward pass with state caching followed by a single backward co-state pass per mode, making the Jacobian cost less than the initial frequency sweep, which is what puts the inverse problem within interactive reach in a browser.

Second, hard equality frequency constraints --- locking selected modes while reshaping others --- can be enforced to first order via the saddle-point formulation in profile-radius space. The large null-space ($N=200$ DOF, $m\leq12$ active modes) makes the linear system well conditioned regardless of the Tikhonov parameter $\lambda$. Nonlinear residuals are reduced by an iterative correction loop; when the loop does not converge the app provides an explicit warning and prompts the user to apply and re-propose incrementally. 

Third, offering the user a family of seven solutions along a logarithmically spaced $\lambda$ sweep, rather than a single automated result, transfers physically meaningful design choices --- how much bore change is acceptable to reach a spectral target --- from the algorithm to the maker.

The direct model was validated against COMSOL FEM at 0.45 cent RMS without wall losses and 1.21 cent with them, over eleven geometries and 132 mode pairs, and below 0.5 cent above the second mode of the six commercial bores. Experimental tap-tone validation against sixteen NC-machined didgeridoos spanning six bore designs yields 9.6 cent RMS over 189 matched mode pairs. 

Regarding experimental validation, because four of the six validation classes are represented by several specimens milled from one CAD file, the reproducibility of the instruments themselves could be measured independently of any model: nominally identical didgeridoos differ from one another by $\sigma_\text{fab} = 4.5$ cent per resonance, with no specimen measurably sharp or flat as a whole.

Exploiting this knowledge we experimentally found that about 36\% of the residual variance is the fabrication scatter quantified above, leaving 6.0 cent for geometry error and model deficiency combined, five times the numerical error of the solver. The inversion validation asks for bore corrections of 0.36 mm RMS pooled; where a within-class comparison is possible these corrections are, class by class, only between $0.74$ and $1.21$ times the bore difference between two physical copies of the same design, i.e. comparable to the spread between copies. We conclude that geometry uncertainty accounts for much, though not all, of the discrepancy. The exception is the widest-belled bore, which hits the limits of the plane-wave TMM as formulated here.

Finally, as an application note, we conclude that \tl{} is well suited to didgeridoo makers' everyday workflow, specifically addressing their needs both in terms of choice of musical targets and constraints and in terms of ease of use. 

\subsection*{Author Contributions}

C. A. R. designed the methodology, wrote and maintained the software, analyzed the data and wrote the original draft. A. F. designed the bore profiles and built the sample instruments, performed the impulse response and sound measurements and tested the software. Both authors revised and edited the draft.

\bibliographystyle{plainnat}
\bibliography{biblio}

\end{document}